\documentclass[aip,author-numerical,jap,preprint]{revtex4-1}

\usepackage{graphicx}
\usepackage{dcolumn}
\usepackage{bm}
\usepackage{times}
\usepackage{color}

\begin{document}
\title{Photo-Seebeck effect in tetragonal PbO single crystals}

\affiliation{Department of Physics, Nagoya University, Nagoya 464-8602, Japan}
\author{P.S. Mondal}
\author{R. Okazaki}
\author{H. Taniguchi}
\author{I. Terasaki}

\begin{abstract}

We report the observation of photo-Seebeck effect in tetragonal PbO crystals. 
The photo-induced carriers contribute to the transport phenomena, 
and consequently the electrical conductivity increases and the Seebeck coefficient decreases with increasing photon flux density. 
A parallel-circuit model is used to evaluate the actual contributions of 
photo-excited carriers from the measured transport data. 
The photo-induced carrier concentration estimated from the Seebeck coefficient increases almost linearly with increasing photon flux density, 
indicating a successful photo-doping effect on the thermoelectric property.
The mobility decreases by illumination but the reduction rate strongly depends on the illuminated photon energy. 
Possible mechanisms of such photon-energy-dependent mobility are discussed.

\end{abstract}

\pacs{72.20.Pa,72.40.+w}
\maketitle

\section{Introduction}

Thermoelectric power generation is a simple and
environmentally-friendly energy-conversion technology.\cite{CRC_1996,PT_42_42_1997,SSP_51_81_1998,NM_7_105_2008}
One of the most important features in thermoelectrics is that it can be adapted in various situations owing to its simple structure without any moving parts compared with other heat engines. This is highly favorable for harvesting electricity from waste heat, a large amount of which is exhausted from a wide variety of heat sources such as automobiles or power plant boilers.

It has recently been shown that the solar light is also used as the heat source to effectively produce the electricity through a thermoelectric device.\cite{NM_10_532_2011,NL_10_562_2010}
This utilizes the solar thermal process, known as the solar thermoelectric generator. Another way for a usage of the solar light with thermoelectrics is to employ the photovoltaic effect. When the photon energy of irradiated light exceeds the band-gap energy, the electron-hole pairs are created by illumination in insulating materials. Such excited carriers should contribute to the electrical transport through changing carrier concentration and/or mobility with illumination. Therefore, the photo-Sebeeck effect, the Seebeck effect of photo-induced carriers, is expected to be observed under illumination along with well-known photoconduction phenomena. 

The photo-Seebeck effect in conventional semiconductors has been reported in earlier studies \cite{CJP_5_528_1955,JAP_41_3182_1970,JAP_41_765_1970,JAP_44_138_1973} and recently also demonstrated in an oxide semiconductor.\cite{JPSJ_81_114722_2012} In general, the Seebeck coefficient $S$ decreases with increasing electrical conductivity $\sigma$ under the light illumination as a consequence of photo-doping effect. Thus the illumination provides a new possible route to improve the efficiency of thermoelectrics, which has been tuned through the chemical doping traditionally.\cite{SSP_51_81_1998} More interestingly, a nontrivial behavior of photon-induced carriers has also been seen in $p$-type silicon.\cite{JAP_41_765_1970} Near room temperature, both the Seebeck coefficient and the conductivity increase with illumination, which is difficult to be understood within the photo-doping effect. 
This may serve as a possible way to enhance the power factor $S^2\sigma$ using photon irradiation.

Lead monoxide has been used as a photoconductive layer in a camera tube.\cite{PRRS_1961} Both the polymorphs of the lead monoxide, tetragonal ($\alpha$-PbO) and orthorhombic ($\beta $-PbO) phases, are known to be photoconductive and having indirect band gaps of $E_g=$ 1.9 and 2.7 eV, respectively.\cite{PRRS_1961,JAP_37_3505_1966,JCG_47_568_1979} In this study, we demonstrate the photo-Seebeck effect in $\alpha$-PbO single crystals as functions of photon energy and photon flux density. We find that the electrical conductivity increases and the Seebeck coefficient decreases with increasing photon flux density. This effect is explained in terms of the photo-doping effect in an intrinsic semiconductor. A parallel-circuit model is used to evaluate the actual contributions of photo-excited carries from the measured transport data. The photo-induced carrier concentration and mobility are evaluated, and the mobility is found to decrease in a photon-energy-dependent manner with increasing photon flux density.

\section{Experimental}

The $\alpha$-PbO single crystals were made with a melt-grown method. 
An alumina crucible containing PbO (5N) powder was heated up to 1163 K in an electrical furnace with a heating rate of 100 K/h. 
After confirming the complete melting of PbO powder, 
the furnace was cooled down slowly with a cooling rate of 2 K/h, 
and at 1133 K the electrical power of the furnace was switched off. 
The sufficiently large and lamellar $\alpha$-PbO single crystals with a typical size of 2 $\times$ 1 $\times$ 0.2 mm$^3$ were obtained.
The band-gap energy was estimated to be 1.8(1) eV from the reflectivity measurement. The conductivity and the Seebeck coefficient were measured at 300 K as functions of photon energy and photon flux density with a home-made experimental setup described elsewhere.\cite{JPSJ_81_114722_2012} The photon energy was varied by selecting light emitting diodes (LEDs) with different photon energies, while the photon flux density was changed by varying LED excitation current. The photon flux density of the incident light was obtained from the illuminance measured using an illuminance meter.

\section{Results and discussions}

Owing to the very low conductivity of the order of $10^{-8}$ $\Omega^{-1}$cm$^{-1}$ at room temperature, the conventional open-circuit voltage measurement technique was not available to measure the Seebeck coefficient. Instead, we employed a current measurement method by exploiting transport equation $J = \sigma E + \sigma S (-\nabla T)$, where $J$, $E$, and $\nabla T$ are the current density, electric field, and temperature gradient, respectively. We first measured the current $I$ in a dc voltage $V$ under no temperature gradient ($\nabla T=0$), and obtained the conductivity $\sigma = J/E = (I/V)(l/A)$, where $l$ and $A$ are the electrode distance and cross-sectional area of the sample, respectively. Then we measured the current under $\nabla T$ without voltage ($V=0$). To eliminate the offset contributions to the photo-Seebeck signal, we measured the current under different $\nabla T$. Figure 1 depicts $\nabla T$ variations of $J/\sigma$ measured in different photon flux densities $P$. The Seebeck coefficient at each photon flux density was obtained from the slope in $J/\sigma$-$\nabla T$ curves.

\begin{figure}[t!]
\centering
\includegraphics[width=0.5\linewidth]{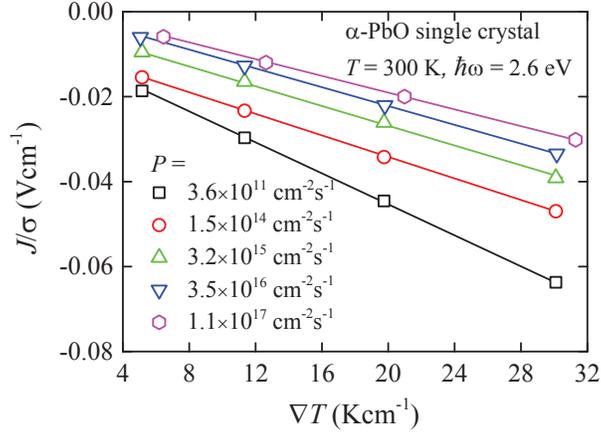}
\caption{ 
(color online) The measured current density divided by the conductivity $J/\sigma$ as a function of temperature gradient $\nabla T$ for different photon flux densities $P$, which are controlled by varying LED excitation current. The photon energy is fixed to be $\hbar\omega$ = 2.6 eV.
The solid lines are linear fitting results.
}
\end{figure}

In Fig. 2(a), we show the electrical conductivity of $\alpha$-PbO as a function of photon flux density measured with different photon energies, $\hbar\omega$ = 1.9, 2.6, and 3.4 eV. While the conductivity is very low at dark ($P \simeq 10^{11}$ cm$^{-2}$s$^{-1}$), it significantly increases with light illumination with photon energies larger than the band-gap energy of $\alpha$-PbO. Note that the photoconduction signal was insignificant for a light with photon energy of $\hbar\omega$ = 1.5 eV (not shown), indicating a contribution of photo-induced carriers across the band gap to the transport phenomena. As seen in Fig. 2(a), the photon flux density variations of $\sigma$ are observed to obey a power law of the form $\sigma \propto P^{y}$ irrespective of photon energies. 
The value of the exponent $y$ is found to be $\sim$ 0.4, which suggests that the recombination process is close to a bimolecular type.\cite{CPAP_1963}

\begin{figure}[t!]
\centering
\includegraphics[width=0.5\linewidth]{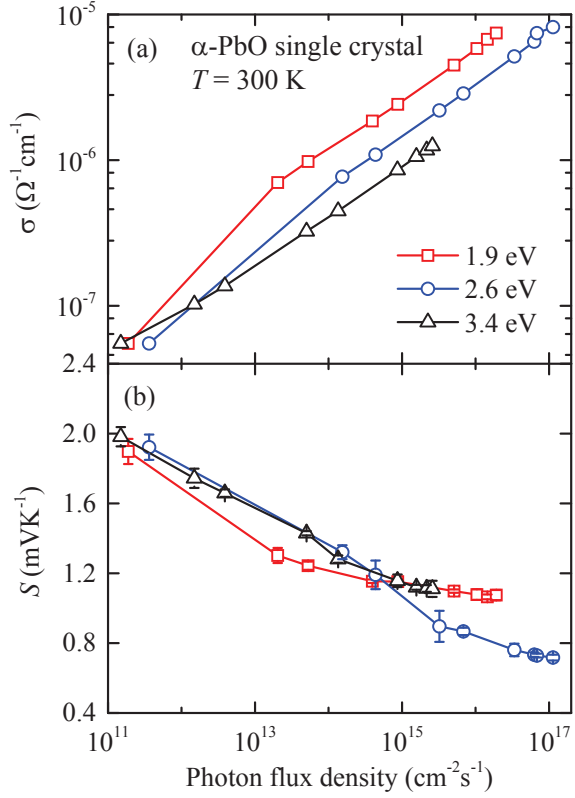}
\caption{(color online) (a) The electrical conductivity and (b) Seebeck coefficient of $\alpha$-PbO as a function of photon flux density measured under the light illuminations with photon energies $\hbar \omega$ = 1.9, 2.6, and 3.4 eV.
}
\end{figure}

Figure 2(b) shows the photo response of the Seebeck coefficient. 
A fairly large positive value of $S$ ($\sim$ 2 mV/K) is observed for $\alpha$-PbO crystals at dark, indicating an intrinsic $p$-type conduction. Under illumination, the value of $S$ systematically decreases with increasing photon flux density and slightly depends on photon energy.
However, it should be noted that the absorption coefficients of $\alpha$-PbO at $\hbar\omega$ = 1.9, 2.6, and 3.4 eV are 
approximately equal to 1 $\times$ 10$^2$, 2 $\times$ 10$^3$, and 1.5 $\times$ 10$^4$ cm$^{-1}$, respectively.\cite{JAP_37_3505_1966} 
Therefore, the illuminated light can penetrate only into a very thin region near the surface of the bulk crystal,
which significantly varies with photon energy. 
Thus photo transport emerges as a surface-dominated phenomenon which must be taken into account in order to analyze the photon flux density and photon energy dependence of the transport behaviors.

A simple parallel-circuit model composed of thin photo-conducting and thick insulating layers is applied to evaluate the actual photo-induced conductivity ($\sigma_{\rm{light}}$) and Seebeck coefficient ($S_{\rm{light}}$) from the measured $\sigma$ and $S$, which are related by the following equations \cite{JPSJ_81_114722_2012} 
\begin{equation}
\sigma = \left(1-\frac{\lambda}{d}\right)\sigma_{\rm{dark}} + \frac{\lambda}{d}\sigma_{\rm{light}},
\label{1}
\end{equation}                                                            
\begin{equation}
\sigma S = \left(1-\frac{\lambda}{d}\right)\sigma_{\rm{dark}} S_{\rm{dark}} + \frac{\lambda}{d}\sigma_{\rm{light}} S_{\rm{light}},
\label{2}
\end{equation} 
where $d$ is the sample thickness and $\lambda$ is the absorption length of light. 
For the bulk crystals ($d\gg\lambda$), $\sigma_{\rm{light}}$ and $S_{\rm{light}}$ approximately take the forms of
\begin{equation}
\sigma_{\rm{light}} = \frac{d}{\lambda}(\sigma-\sigma_{\rm{dark}}),
\label{3}
\end{equation}  
\begin{equation}
S_{\rm{light}} = \frac{\sigma S - \sigma_{\rm{dark}} S_{\rm{dark}}} {\sigma - \sigma_{\rm{dark}}}.
\label{4}
\end{equation} 

In Fig. 3, we show the relationship between $\sigma_{\rm{light}}$ and $S_{\rm{light}}$ under the light illuminations with different photon energies. 
The $S_{\rm{light}}$-$\ln\sigma_{\rm{light}}$ curves roughly exhibit linear relationships with photon-energy-dependent slopes
which are found to be larger than $k_B/e$. The result indicates that the photo-induced carrier mobilities are appreciably different with photon energies and are not constant throughout the photo-doping process for a fixed photon energy.

\begin{figure}[t!]
\centering
\includegraphics[width=0.5\linewidth]{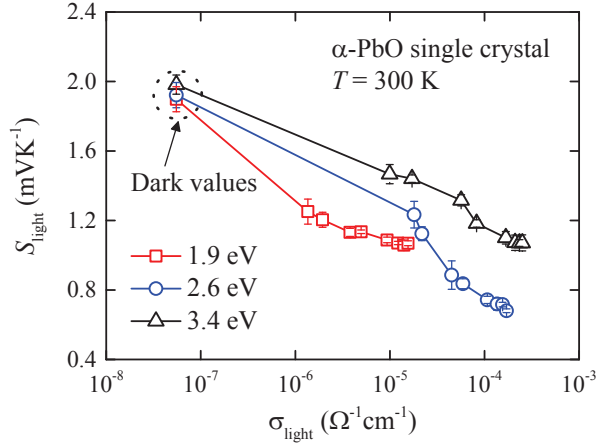}
\caption{(color online) The relation between light induced conductivity ($\sigma_{\rm{light}}$) and Seebeck coefficient ($S_{\rm{light}}$) in $\alpha $-PbO single crystal under the light illuminations with photon energies $\hbar \omega$ = 1.9, 2.6, and 3.4 eV.
The Seebeck coefficient and the conductivity measured at dark are also plotted as indicated by the dashed circle.}
\end{figure}

In the following we will discuss the photo-induced carrier concentration change on the basis of conventional semiconductor physics, in which the Seebeck coefficient is given as\cite{PRB_61_5303_2000}        
\begin{equation}
S_{\rm{light}} = \frac{k_B}{q}[-y+\delta_r(y)],
\label{5}
\end{equation}
where                                                  
\begin{equation}
\delta_r(y) = \frac{(r+2)F_{r+1}(y)}{(r+1)F_{r}(y)}.
\label{6}
\end{equation}
Here \textit{q} is the charge of the carriers, 
\textit{r} is the scattering parameter ($r = - 1/2$ for electron-phonon scattering and 3/2 for impurity scattering), 
$y=\mu/k_B T$ is the reduced chemical potential, and $F_r(y)$ is the Fermi integral 
expressed as
\begin{equation}
F_r(y) = \int_{0}^{\infty }\frac{x^r}{1+e^{x-y}}dx.
\label{7}
\end{equation}
We now calculate the value of $F_r(y)$ from the $S_{\rm light}$ by assuming $r = 3/2$, and 
then estimate the photo-induced carrier concentration given as\cite{PT_42_42_1997,SSP_51_81_1998}
\begin{equation}
n = z \left(\frac{2\pi m_h^* k_B T}{h^2}\right)^{3/2} \frac{2}{\sqrt{\pi}}F_{1/2}(y),
\label{8}
\end{equation}                                                              
where $m_h^*$ is the effective mass of hole which is taken to be $m_h^*= 2.44 m_0$ 
($m_0$ is the bare mass of electron)\cite{EPL_99_47005_2012} 
and $z$ is the degeneracy of the band.
The valence band of $\alpha$-PbO mainly consists of six non-degenerate bands 
which are comprised of the hybridization between O $2p$ and Pb $6s$ orbitals. 
Therefore, including spin degeneracy one can roughly consider \textit{z} = 2.

\begin{figure}[t!]
\begin{center}
\includegraphics[scale=0.39]{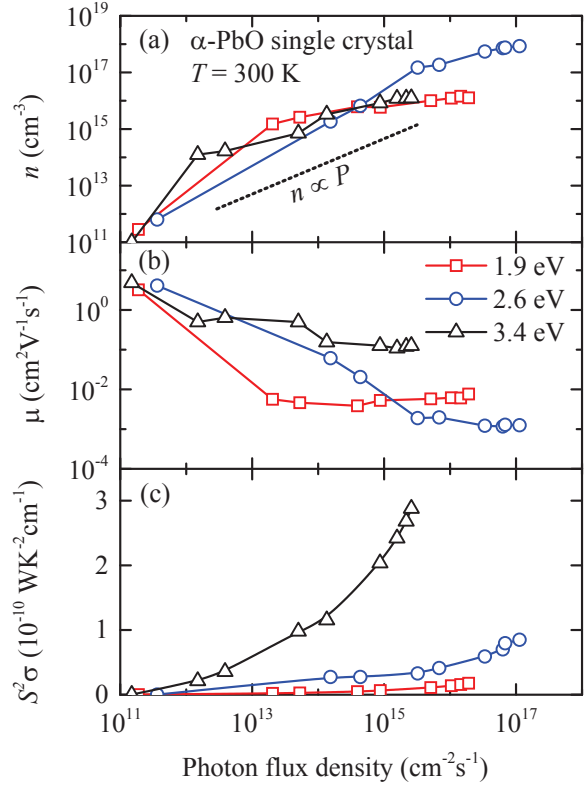}
\caption{(color online) Photo-induced (a) carrier concentration $n$, (b) carrier mobility $\mu$, and (c) power factor $S^2\sigma$ as a function of photon flux density under the light illuminations with photon energies $\hbar \omega$ = 1.9, 2.6, and 3.4 eV. The dashed line in (a) shows a linear relation between $n$ and $P$.
}
\end{center}
\end{figure}

In Fig. 4(a), we show the photon flux density dependence of carrier concentration $n$ of $\alpha$-PbO for photon energies $\hbar\omega$ = 1.9, 2.6, and 3.4 eV. 
The typical value of dark carrier concentration is of the order of  $10^{11}$ cm$^{-3}$, which is comparable to that
of intrinsic semiconductors. 
Under illumination, the carrier concentration almost linearly increases with increasing photon flux density ($n\propto P$)
by several orders of magnitude
with slight photon energy dependence,
indicating a successful photo-doping into the crystal surface region.

We then estimate the mobility of photo-induced carriers 
by using the relation $\mu = \sigma_{\rm {light}}$/$qn$. 
The carrier mobility at dark is found to be $\sim$ 5 cm$^{2}$V$^{-1}$s$^{-1}$.
This value is smaller than that reported in the Hall measurements on $\alpha$-PbO crystals ($\mu\sim$ 50 cm$^{2}$V$^{-1}$s$^{-1}$),\cite{JAP_39_2062_1968} 
probably due to inevitable crystalline impurities.
In Fig. 4(b), the photo-induced mobility change for three photon energies are shown as a function of photon flux density. 
The mobility shows a remarkable photon energy dependence 
in high contrast to the carrier concentration behavior. 
As the photon flux density increases, a step like decrease of mobility by two orders of magnitude from the dark value followed by a saturation behavior is found for the light illuminations with $\hbar\omega$ = 1.9, 2.6 eV. However, for $\hbar\omega$ =  2.6 eV, 
the saturation behavior occurs at a higher photon flux density than that for $\hbar\omega$ = 1.9 eV. 
In contrast, for $\hbar\omega$ = 3.4 eV, 
the mobility gradually reduces with increasing photon flux density.

In Fig. 4(c), the photon flux density variations of power factor $S^2\sigma$ for different photon energies are plotted. 
The power factor progressively increases with increasing photon flux density. 
For $\hbar\omega$ = 3.4 eV, 
the largest photo-induced increase of power factor is observed which is attributed to the fact that 
the photo-excited carriers with $\hbar\omega$ = 3.4 eV exhibit higher mobility than that with other photon energies. 
Nevertheless, the optimum value of $S^2\sigma$ was not achieved 
by illuminations.

We discuss an origin of the photon-energy-dependent mobility.
In $\alpha$-PbO,
depending on the crystal-growth conditions, 
the presence of lead and/or oxygen vacancies are unavoidable, 
and non-compensated ionized vacancies can act as charged scattering centers.\cite{jpcm_25_075803_2013} 
The photo-excitation is expected to increase the density of charged scattering centers 
through changing the effective charge of vacancies by capturing photo-excited carriers.\cite{PRB_68_064101_2003} 
Since the mobility is inversely proportional to the density of scattering centers, 
the photo-excitation can appreciably decrease the mobility.\cite{PRB_121_473_1961} 
Moreover, in such an ionized-impurity scattering regime,
the mobility also depends on the electron energy $\epsilon$
through the energy-dependent scattering time $\tau(\epsilon)\propto\epsilon^{3/2}$.\cite{RMP_53_745_1981}
Thus the photo excitation with higher photon energy leads to carriers with larger scattering time (higher mobility).
The observed photon-energy-dependent mobility is qualitatively explained within this scattering regime.

Another possible origin of the strong photon-energy dependence of the relation between the photo-Seebeck effect and the photoconduction (Fig. 3) is 
a photon-energy-dependent effective mass for the photo-transport behaviors.
Now $\alpha$-PbO is an indirect semiconductor having a non-degenerate dispersive conduction band at M point 
and a valence band consisting of six non-degenerate bands at $\Gamma$ point.\cite{EPL_99_47005_2012} 
It seems that photo-excitation with photon energy comparable to the band-gap energy can create heavy holes on the top the valence band,
while higher-energy photons yield lighter holes in the deep of the valence band.
Such a difference of the effective mass with the photon energy 
possibly leads to 
the photon-energy dependence of the photo-excited carrier transport.
To examine the detailed photon-energy variations of the scattering time and the effective mass,
the photo-Hall measurement is required as a future study.

\section{Conclusions}
In summary, we have successfully observed the photo-Seebeck effect in $\alpha$- PbO single crystal. 
The illumination significantly increases the electrical conductivity and decreases the Seebeck coefficient. 
The actual contributions of the photo-induced carriers to the transport are extracted 
by using a parallel-circuit model. 
The photo-induced carrier concentration increases in a linear proportion with the photon flux density and 
the carrier mobility decreases in accompanying with remarkable photon-energy dependence. 
We suggest that the charged impurity scattering is responsible for the energy-dependent decrease of the mobility. 
The power factor is progressively increased with photon intensity as a consequence of photo-doping effect. 
However, the photon-irradiation is found to be insufficient to achieve the optimal carrier concentration in $\alpha$- PbO.

\section*{ACKNOWLEDGMENTS}

This work was supported by Advanced Low Carbon Research and Development Program (ALCA) from Japan Science and Technology Agency (JST), Program for Leading Graduate Schools ``Integrative Graduate Education and Research Program in Green Natural Sciences (IGER)'' from Ministry of Education, Culture, Sports, Science and Technology (MEXT), 
and the Mitsubishi Foundation.


\end{document}